\documentclass[11pt]{article}
\usepackage{epsf,amsmath,amsfonts,times,cite}

\newcommand{\resection}[1]{\setcounter{equation}{0}\section{#1}}
\newcommand{\One}{{\hbox{{\rm 1{\hbox to 1.5pt{\hss\rm1}}}}}}

\newcommand{\bc}[6]{{\,C^{(\!#1\hspace{-0.3pt}#2\hspace{-0.3pt
     }#3\!)#6}_{\;#4#5 }\,}}

\newcommand{\eps}{\varepsilon}
\newcommand{\F}[6]{{\ensuremath{\textsf{\large F}^{}_{#5#6}\!\left[
     \begin{array}{ll}
     \!#2 & \!#3\! \\ \!#1 & \!#4\! 
     \end{array}\!\right]}}}
\newcommand{\lam}{\lambda}

\newcommand{\mZ}{\mathbb{Z}}

\newcommand{\blank}[1]{{}}

\newcommand{\con}[1]{#1^\vee}
\newcommand{\il}{{\langle \! \langle}}
\newcommand{\ir}{{\rangle \! \rangle}}
\newcommand{\tr}{{\text{Tr}}}

\renewcommand{\ss}{\scriptstyle}
\newcommand{\mb}[2]{\makebox(0,0)[#1]{$\ss{#2}$}}

\begin{document}

\begin{titlepage}
\vskip 0.5cm
\begin{flushright}
LPTHE-P03-08\\
KCL-MTH-03-05\\
{\tt hep-th/0306167}\\
\end{flushright}
\vskip 4.cm
\begin{center}
{\Large\bf Defect Lines and Boundary Flows } \\[5pt]
\end{center}
\vskip 1.3cm
\centerline{16th June 2003}
\vskip 0.6cm
\centerline{K.~Graham\footnote{e-mail: {\tt kgraham@lpthe.jussieu.fr}}
and G.M.T.~Watts\footnote{e-mail: {\tt gmtw@mth.kcl.ac.uk}}
}
\vskip 0.6cm
\centerline{${}^1$\sl LPTHE, Universit\'e Paris VI,}
\centerline{\sl 4 place Jussieu, F\,--\,75\,252\, Paris\, Cedex 05, France}
\vskip 0.4cm
\centerline{${}^2$\sl Mathematics Department, }
\centerline{\sl King's College London, Strand, London WC2R 2LS, U.K.}
\vskip 0.9cm
\begin{abstract}
\vskip0.15cm
\noindent
Using the properties of defect lines, we study boundary
renormalisation group flows.  We find that when there exists a flow
between maximally symmetric boundary conditions $a$ and $b$ then there
also exists a 
boundary flow between $c \times a$ and $c \times b$ where $\times$
denotes the fusion product.
We also discuss applications of this simple observation.
\end{abstract}
\end{titlepage}
% The title

\resection{Introduction}
\label{sec:i}

Conformal field theories describe statistical systems at criticality.
In two dimensions, the classification of classes of conformal field
theory has provided an understanding of the notion of universality.
When one introduces a boundary to the system, new elements arise and
one may attempt to classify boundary critical phenomena.

The machinery of rational conformal field theory is well developed and
a great deal is known about the systems at criticality.  Most less
well understood are theories away from the critical point.  A useful
way to study such theories is to deform a well known conformal field
theory by some relevant perturbation.  Using the knowledge of the
original theory one may study the perturbed theory using a variety of
techniques:  Perturbation theory 
%\cite{RecRogSch00,FreSch02,Gra02}
and the truncated conformal space
%\cite{YurZam90,DorPocTatWat98,DorRunTatWat00,GraRunWat00}
approach being two.  If the perturbation is also integrable one can
also use a variety of 
powerful results from the theory of integrable systems: for example,
TBA methods
%\cite{LesSalSim98,FevPeaRav02} 
or non-linear integrable equations
%\cite{}
.

In this note, we consider boundary perturbations within rational
conformal field theory.  Using the technology of defect lines we
observe a useful embedding of operator algebras of the boundary
condition $a$ into the operator algebra of boundary $a \times b$ for
some $b$ (where in the cases we will consider, boundary conditions are
labelled by representations of the chiral algebra and $\times$ denotes
the fusion product of 
these representations).  This embedding
is used to show the main result of this paper. \\

{\em \noindent Assume there exists a boundary flow from boundary condition
  $b$ to $d$, then there also exists a flow from boundary condition
  $a\times b$ to $a \times d$. } \\

Using this result, we prove the existence of many flows
conjectured in the literature using the proven existence of some
elementary flows.  We also consider the conjecture of Fredenhagen and
Schomerus \cite{FreSch02a,Fre03}.  Here we show that a subset of the
flows predicted by their conjecture may be obtained 
from the knowledge of a smaller number of elementary flows. \\

An outline of the paper is as follows.  In section \ref{sec:intro} we give
a lightning review of boundary rational conformal field theory, to set
notation.  Here we also review the construction of defect lines
following Petkova and Zuber.  To study boundary perturbations we will
use the formalism of TCSA.  Note that we will not do any numerical
analysis here, all our results are exact, however the formalism of
TCSA provides a powerful tool to study these deformed boundary
theories.  In section \ref{sec:t} we introduce these ideas.
Using all this technology, we prove our theorem in section
\ref{sec:dlandbs} before demonstrating numerous applications in section
\ref{sec:apps}.  We end with some conclusions and directions for further work.

\resection{A Lightning Review of Rational Conformal Field Theory}
\label{sec:intro}

We consider a rational conformal field theory (RCFT) built from two copies of
a chiral algebra $\mathcal{A}$ containing the Virasoro algebra, and a finite
set of representations of $\mathcal{A}$ which we denote by $\mathcal{R}_i$,
$i=1,2,\ldots,n$.  

The spectrum of the bulk theory is encoded in the torus partition
function,
\begin{align}
  Z = \sum_{i,\bar{i}} Z_{i,\bar{i}} \, \chi_i(q)
  \bar{\chi}_{ \bar{i} } (\bar{q})  \; .
\end{align}
For simplicity, we consider only the charge conjugation modular
invariant,
\begin{align}
  Z_{i,\bar{i}} = C_{i,\bar{i}} = \delta_{i,\con{\bar{i}}}
\end{align}
where $C$ is the charge conjugation matrix relating a representation
labelled by $i$, to that of its conjugate $\con{i}$.  In our RCFT we
have Verlinde's formula,
\begin{align}
  N_{ij}{}^k = \sum_\ell \frac{ S_{i\ell} S_{j\ell} S_{k\ell}^* }{
  S_{0\ell} } \; .
  \label{eq:verlinde}
\end{align}
Where the S-matrix is symmetric, unitary and satisfies,
\begin{align}
  S^2  =  C \; , \qquad 
%  C_{ij} = \delta_{j\con{i}} \; , \qquad
  S_{ij^\vee} = S_{ij}^* \; .
\end{align}
Under modular transformations, the characters transform as,
\begin{align}
  \chi_i(q) = \sum_j S_{ij} \chi_j(\tilde{q}) \; , \qquad 
  q = e^{ 2 \pi i \tau } \; , \qquad
  \tilde{q} = e^{ - 2 \pi i / \tau } \; .
\end{align} 
In particular, this implies,
\begin{align}
  \chi_j(\tilde{q}) = \sum_i S_{j\con{i}} \chi_i(q) = \sum_i S_{ji}^* \chi_i(q) \; .
\end{align}
The Fusion numbers satisfy the following identities,
\begin{align}
  &N_{ab}{}^c = N_{ba}{}^c \; ,
  \qquad
  \sum_k N_{ab}{}^k N_{kc}{}^d = \sum_\ell N_{ac}{}^\ell N_{\ell
  b}{}^d \; , \\
  &N_{0a}{}^b = \delta_{ab} \; , \qquad N_{ab}{}^0 = \delta_{a\con{b}}
  \; ,
  \qquad
  N_{ab}{}^c = N_{\con{a} \con{b}}{}^{\con{c}} 
\end{align}

Boundary conditions preserving a maximal amount of symmetry are given
by Cardy's construction, which we now review.
Maximally symmetric boundary conditions in RCFT are a restricted class
of objects satisfying \cite{Ish89,Car89},
\begin{align}
  \left( L_n - \bar{L}_{-n} \right) |a \rangle_\Omega = 0 \; , \qquad
  \left( W_n - (-1)^{s_W} \Omega \bar{W}_{-n} \right) | a
  \rangle_\Omega = 0 \; ,
  \label{eq:ishcond}
\end{align}
where $L_n$ are the Virasoro generators, $W_n$ are the modes of a
field from the chiral algebra $\mathcal{A}$ and $s_W$ is its
spin. We have also introduced a gluing automorphism, $\Omega$ 
\cite{KatOka97,RecSch98,FucSch98}. 
These equations may be solved separately in each sector of the bulk
Hilbert space where one finds a unique solution, given by an Ishibashi
state $| i \ir$ with $i$ labelling the sector. 
% General solutions are
%linear combinations of these states that satisfy an extra condition
%called Cardy's condition.  
The Ishibashi states are a basis of the boundary states but do not
correspond to physically realisable boundary conditions. The
physically realisable boundary conditions can be found by imposing an
extra condition, called Cardy's condition.
To formulate this condition, we consider a
cylinder with boundary conditions $\langle a | = \sum_i 
( \psi_a^i )^*  \, \il i |$ on the left and $| b \rangle = \sum_i 
\psi_b^i \, | i \ir $ on the right.  With time running around the
cylinder, the Hilbert space of the theory is,
\begin{align}
  \mathcal{H}_{ab} = \bigoplus_k n_{ka}{}^b \, \mathcal{R}_k \; , 
\end{align}
The partition function for this system is then,
\begin{align}
  Z_{ab} = \sum_k n_{ka}{}^b \chi_k(q) \; ,
  \label{eq:cylpart}
\end{align}
and Cardy's condition is that,
\begin{align}
  \sum_i  \psi_a^i ( \psi_b^i )^* \, S_{ij} = n_{ja}{}^b \; ,
  \label{eq:cardy}
\end{align} 
where $S$ is the modular matrix mentioned above.
A general solution to this equation is not known, the difficulty being
that the numbers $ n_{ja}{}^b$ should be non-negative integers.
However, much is understood \cite{FucRunSch02,BehPeaPetZub99}. 
One important solution was given by Cardy in his original paper
\cite{Car89} for theories with a charge conjugation modular invariant.  
Using the Verlinde formula \eqref{eq:verlinde} one
observes that the ansatz,
\begin{align}
  \psi_a^i = \frac{ S_{ia} }{ \sqrt{ S_{i0} } }
\end{align}
solves Cardy's condition with,
\begin{align}
  n_{jb}{}^a = N_{jb}{}^a  \; .
\end{align}
We will call such boundary conditions, {\em Cardy boundary
conditions}.
These have the property that the identity representation appears once
and once only in the Hilbert space ${\cal H}_{aa}$, so that there is a
unique operator of conformal weight zero on the boundary and the
boundary correlation functions satisfy the physical {\em clustering property}.
  More general boundary conditions may be obtained by
taking linear combinations of these solutions with positive integer
coefficients.  Note that Cardy's 
boundary conditions are also in one to one correspondence with the
primary fields of the bulk theory, and so we label them using the same set.

The spectrum of boundary fields is neatly encoded in the cylinder
partition function \eqref{eq:cylpart}.  A general boundary field
carries five labels: $\phi_{i_\alpha ; {\bf n}}^{(ab)}(x)$, where $i$ labels the
representation, $a$ and $b$ label the boundary conditions on the right
(${}{>}x$) 
and left (${}{<}x$) of the insertion point $x$ respectively, while $\alpha =1,2,\ldots,
N_{ib}{}^a$ accounts for any further multiplicity there may be.  The
label ${\bf n}$ denotes the descendants of the primary field in
representation $i$. % We may suppress this label when discussing
%primary fields :     $\phi_{i_\alpha}^{(ab)}(x) = \phi_{i_\alpha ;
%  {\bf 0}}^{(ab)}(x)$.  
%
%To write the OPE of boundary primary fields we need a little more
%notation.  Let $\{ | i , {\bf n} \rangle \}_{{\bf n}}$ denote an
%orthonormal, $L_0$-graded basis
%for the representation $\mathcal{R}_i$.  We also choose a basis in the
%space of chiral vertex operators, $V_{ij ; \gamma}^{ \phantom{ij;}
%  k}(z) : \mathcal{R}_i \to \text{Hom}(\mathcal{R}_j,\mathcal{R}_k)[[z,z^{-1} ]] \,
%z^{-h_i-h_j+h_k}$ with $\gamma = 1,2,\ldots, N_{ij}{}^k$.  Then the
%OPE has the form,
%\begin{align}
% x>y \; , \qquad \phi_{i_\alpha}^{(ab)}(x) \phi_{j_\beta}^{(cd)}(y) =
%  \sum_{k,\gamma,\delta}  \delta_{bc} \bc acd{i_\alpha}{j_\beta ;
%  \gamma}{k_\delta}
%  \sum_{{\bf n}}
%%  \langle k,{\bf n} | \phi_{i_\delta}(x-y) | j ,{\bf  0} \rangle \,
%  \langle k,{\bf n} | V_{ij ; \gamma}^{ \phantom{ij} k}(x-y) | j ,{\bf  0} \rangle \,
%  \phi_{k_\delta ; {\bf n}}^{(ad)}(y) \; ,
%\end{align}
The OPE has the form,
\begin{align}
 x{>}y \; , \quad \phi_{i_\alpha; {\bf n}}^{(ab)}(x) \phi_{j_\beta ;
  {\bf m}}^{(cd)}(y) =
  \sum_{k,{\bf p}}
  \sum_{\gamma=1}^{N_{ij}{}^k} \sum_{\delta=1}^{N_{kd}{}^a}
  \delta_{bc} \bc acd{i_\alpha}{j_\beta ; \gamma}{k_\delta}
%  \sum_{{\bf p}}
  \, \beta^\gamma \!\! \left[ _{i ; {\bf n} \;j ; {\bf m}}^{\;\;\; k ;
  {\bf p}} \right]
  (x{-}y)^{{-}h_{i;{\bf n}}{-}h_{j;{\bf m}}{+}h_{k;{\bf p}}}
  \phi_{k_\delta ; {\bf p}}^{(ad)}(y) \; ,
  \label{eq:OPE}
\end{align}
where the numbers $\beta^\gamma [ \cdot ]$ are entirely determined
by the chiral algebra $\mathcal{A}$ and the structure constants $C$ satisfy the sewing constraints of 
\cite{CarLew91}. In the case of the Virasoro minimal models, these sewing
constraints have been solved by Runkel \cite{Run99} who showed that
there exists a normalisation in which,
\begin{align}
%   \bc abcijk = \F iacjbk
  \bc abcijk = \F aijckb
  \label{eq:ingossol}
\end{align}
where $\textsf{\large F}$ is the fusing matrix which encodes the associativity of the
operator product expansion.
In more general RCFT, this result generalises
\cite{BehPeaPetZub99,FelFroFucSch99a,FelFroFucSch99b} and there exists a
normalisation in which\footnote{We thank Ingo Runkel for discussions
  on the following.} \cite{FelFroFucSch99b},  
%In the notatio\sum_{\alpha =1}^{N_{ab}{}^k} \sum_{\beta =
%  1}^{N_{ab}{}^\ell} n of \cite{BehPeaPetZub99} we have,
%\begin{gather}
%  \bc abc{i_\alpha}{j_\beta ; \delta}{k_\gamma} = 
%  { \F aijcbk }_{\alpha,\beta}^{\gamma, \delta} \; , \\
%  \alpha = 1,2, \ldots , N_{ib}{}^a \; , \qquad
%  \beta = 1,2, \ldots , N_{jc}{}^b \; , \\
%  \gamma = 1,2, \ldots , N_{kc}{}^a \; , \qquad
%  \delta = 1,2,\ldots , N_{ij}{}^k \; , \\
%\end{gather}
\begin{gather}
  \bc abc{i_\alpha}{j_\beta ; \gamma}{k_\delta} = \left( 
  { \F {\con a}ijck{\con b} }_{\alpha,\beta}^{\gamma, \delta}
  \right)^* \; , \\
  \alpha = 1,2, \ldots , N_{ib}{}^a \; , \qquad
  \beta = 1,2, \ldots , N_{jc}{}^b \; , \\
  \gamma = 1,2, \ldots , N_{ij}{}^k \; , \qquad
  \delta = 1,2,\ldots , N_{kc}{}^a \; , 
\end{gather}
where the $\mathsf{F}$-matrix relates different bases for four point
conformal blocks (see \cite{MooSei89} for a definition),
\begin{align}
\begin{array}{c} \mbox{
\setlength{\unitlength}{1.5pt}
  \begin{picture}(20,35)
    \put(10,10){\line(1,-1){10}}
    \put(10,10){\line(-1,-1){10}}
    \put(10,10){\line(0,1){15}}
    \put(10,25){\line(1,1){10}}
    \put(10,25){\line(-1,1){10}}
    \put(-0.5,-0.5){\mb{tr}{i}} 
    \put(20.5,-0.5){\mb{tl}{\ell}} 
    \put(-0.5,35.5){\mb{br}{j}} 
    \put(20.5,35.5){\mb{bl}{k}} 
    \put(7.5,17.5){\mb{r}{p}} 
    \put(10.5,24.5){\mb{lt}{\gamma}} 
    \put(10.5,10.5){\mb{lb}{\delta}} 
    \put(10,17.5){\vector(0,-1){1}}
    \put(5,5){\vector(1,1){1}}
    \put(15,5){\vector(-1,1){1}}
    \put(5,30){\vector(1,-1){1}}
    \put(15,30){\vector(-1,-1){1}}
  \end{picture}  
} 
\end{array}
= \sum_{k,\alpha,\beta} { \F ijk{\ell}pq }_{\alpha,\beta}^{\gamma,\delta}
\begin{array}{c} \mbox{
\setlength{\unitlength}{1.5pt}
  \begin{picture}(35,20)
    \put(10,10){\line(-1,1){10}}
    \put(10,10){\line(-1,-1){10}}
    \put(10,10){\line(1,0){15}}
    \put(25,10){\line(1,1){10}}
    \put(25,10){\line(1,-1){10}}
    \put(-0.5,-0.5){\mb{tr}{i}} 
    \put(35.5,-0.5){\mb{tl}{\ell}} 
    \put(-0.5,20.5){\mb{br}{j}} 
    \put(35.5,20.5){\mb{bl}{k}} 
    \put(17.5,9.5){\mb{t}{q}} 
    \put(25.5,10.5){\mb{rb}{\beta}} 
    \put(10.5,11.5){\mb{lb}{\alpha}} 
    \put(17.5,10){\vector(1,0){1}}
    \put(5,5){\vector(1,1){1}}
    \put(30,5){\vector(-1,1){1}}
    \put(5,15){\vector(1,-1){1}}
    \put(30,15){\vector(-1,-1){1}}
  \end{picture}  
}
\end{array}
\end{align}
One can check that when one of the legs is replaced by the identity
representation, the matrix degenerates and we find,
\begin{align}
%  { \F aijcbk }_{\alpha,\beta}^{\delta, t}
%  &={ \F {j^\vee}c{a^\vee}i{b^\vee}{k^\vee} }_{\beta,
%  \alpha}^{t,\delta} \; , \qquad &&\text{ 
%  equation $(E.1)$ of \cite{BehPeaPetZub99}.}  \\
 { \F ijk0pq }_{\alpha,\beta}^{\gamma, \delta}
  &= 
  \begin{cases}
    \delta_{p\con{i}} \, \delta_{q\con{k}} \, \delta_{\beta 1} \, \delta_{\delta 1} \, 
  \delta_{\alpha \gamma} \; , \qquad \text{if $N_{ij}{}^{\con{k}}>0$} \, , \\
    0  \; , \qquad \qquad \qquad \quad \qquad \; \; \text{otherwise} \, .
  \end{cases} \; .
% \quad && \text{
%  equation $(4.12)$ of \cite{BehPeaPetZub99}.}
\end{align}
We then have,
\begin{align}
  \bc ab0{i_\alpha}{j_\beta ; \gamma}{k_\delta} 
  =  
  \begin{cases}
    \delta_{ka} \, \delta_{jb} \, \delta_{\beta 1} \, \delta_{\delta
  1} \delta_{\alpha \gamma} \, 
  \; , \qquad 
%  \text{if $N_{\con{a}i}{}^{\con{b}}>0$} \, , \\
  \text{if $N_{ib}{}^{a}>0$} \, , \\
    0  \; , \qquad \qquad \qquad \quad \quad \;\; \text{otherwise} \, .
  \end{cases} \; ,  \label{eq:structurezero}
%  \begin{cases}
%    \delta_{ka} \, \delta_{jb} \, \delta_{\beta 1} \, \delta_{\gamma 1} \, 
%  \; , \qquad \text{if $N_{ib}{}^a>0$} \, , \\
%    0  \; , \qquad \qquad \qquad \quad \; \text{otherwise} \, .
%  \end{cases} \; ,  \label{eq:structurezero}
\end{align}
which we will require later.

\subsection*{Petkova and Zuber's Defect Lines}
\label{sec:defect}

The construction of defect lines is very analogous to that of boundary
conditions. 
Following Zuber and Petkova \cite{PetZub00},
we will define disorder lines as operators $X$, satisfying the
relations,
\begin{align}
  [L_n,X]=[\bar{L}_n,X]=0 \; , \label{eq:disorderdefn1}\\
  [W_n, X]=[ \bar{W}_n,X ] = 0 \; .
  \label{eq:disorderdefn}
\end{align}
One may also envisage putting a gluing automorphism into one or both
of the second relations.
As an operator in the bulk Hilbert space, the operator $X$ is
naturally associated to some cycle.  For example on the plane, this
would be some contractible cycle around the origin.  On the cylinder
this cycle is non-contractible.  The definition
\eqref{eq:disorderdefn} implies that this operator is invariant under
local diffeomorphisms - i.e. the operator $X$ is invariant under
distortions of this line.  Thus we may associate $X$ to the homotopy
class of the cycle to which it is associated.

As in the boundary case and Cardy's condition, there are also a number
of consistency conditions which must be satisfied by the operator
$X$.  To formulate these conditions, one first notes that as a
consequence of (\ref{eq:disorderdefn1},\ref{eq:disorderdefn}), $X$ is a sum of projectors, 
\begin{align}
  X = \sum_{j,\bar{j},\alpha,\alpha'} \Psi^{(j,\bar{j};\alpha,\alpha')}
   P^{(j,\bar{j};\alpha,\alpha')} \; ,
  \label{eq:disorderform}
\end{align}
where $\alpha,\alpha' = 1,2, \ldots,Z_{j,\bar{j}}$ allow for repeated
representations in the Hilbert space and if $\{ | i, {\bf n} \rangle \otimes | \bar{i} , {\bf \bar{n}}
\rangle \}$ denotes an orthonormal basis in $\mathcal{R}_i \otimes
\mathcal{\bar{R}}_{\bar{i}}$, we write,
\begin{align}
  P^{(i,\bar{i};\alpha,\alpha')} 
  = \sum_{{\bf n},{\bf \bar{n}}} 
  (| i, {\bf n} \rangle \otimes | \bar{i} , {\bf \bar{n}} \rangle )^{(\alpha)}
  (  \langle  i , {\bf n} | \otimes \langle  \bar{i} , 
  {\bf \bar{n}} | )^{(\alpha')} \; .
\end{align}
We will also require the projectors to be Hermitian, 
\begin{align}
  \left( P^{(i,\bar{i};\alpha,\alpha')} \right)^\dagger =
  P^{(i,\bar{i};\alpha,\alpha')} \; .  
\end{align} 
We interpret the defect line $X^\dagger$ as the line $X$ but with
opposite orientation.

A consistency condition is found by considering a pair of defect
lines wrapping a canonical cycle on a torus.  Using a Hamiltonian
picture with ``time'' moving perpendicular to the lines, the torus
partition function may be written,
\begin{align}
  Z_{x|y} &\equiv  \tr_{\mathcal{H}} \left( X_x^\dagger X_y \tilde{q}^{L_0-\tfrac{c}{24}}
  \tilde{\bar{q}}^{\bar{L}_0-\tfrac{c}{24}} \right) \; , \\
  &= \sum_{j,\bar{j},\alpha,\alpha'} 
  \left( \Psi_x^{(j,\bar{j};\alpha,\alpha')} \right)^*
  \Psi_y^{(j,\bar{j}; \alpha,\alpha')}
  \chi_j(\tilde{q}) \chi_{\bar{j}}(\tilde{\bar{q}}) \; .
\end{align}
A second representation of the same partition function may be obtained
by considering time running parallel to the defect lines.  In this
case, the definition of the disorder line 
\eqref{eq:disorderdefn} insures one may still construct two sets of
generators $L_n$ and $\bar{L}_n$ satisfying the Virasoro algebra (or
more generally the chiral algebra $\mathcal{A}$).
Hence the Hilbert space decomposes into irreducible representations,
\begin{align}
  \mathcal{H}_{x|y} = \bigoplus_{i,\bar{i}} V_{i \bar{i} ; x}{}^y
  \mathcal{R}_i \otimes \mathcal{\bar{R}}_{\bar{i}} \; ,
\end{align}
for some non-negative integers $V_{i \bar{i} ; x}{}^y$, and the partition
function becomes,
\begin{align}
  Z_{x|y} = \tr_{\mathcal{H}_{x | y} } q^{L_0-\tfrac{c}{24}}
  q^{\bar{L}_0-\tfrac{c}{24}} = \sum_{i,\bar{i}} V_{i \bar{i} ;
  x}{}^y \chi_i(q) \chi_{\bar{i}}(\bar{q}) \; .
\end{align}
We may equate these two expressions using the modular transformation
properties of the characters,
\begin{align}
  V_{i \bar{i} ; x}{}^y = \sum_{j, \bar{j},\alpha,\alpha'} S_{ji}
  S_{\bar{j} \bar{i}} \, \Psi_x^{(j ,
  \bar{j};\alpha,\alpha')\, *} \Psi_y^{(j , 
  \bar{j};\alpha,\alpha')}{} \; .
  \label{eq:ZPconstraint}
\end{align}
The task now is to solve this equation.  A general discussion of
this problem would take us too far a field\footnote{We refer the
 interested reader to \cite{PetZub01,FucRunSch02}.} so here we simply state a
class of solutions to \eqref{eq:ZPconstraint} for theories with a
charge conjugation modular invariant 
and refer the reader to the original paper of Zuber and Petkova
\cite{PetZub00}.   Following \cite{PetZub00} we observe using the
Verlinde formula 
\eqref{eq:verlinde} that the ansatz,
\begin{align}
  \Psi_x^{(i,\bar{i})} = \frac{ S_{xi} }{ S_{0i} } \; ,
  \label{eq:disordersoln}
\end{align}
satisfies \eqref{eq:ZPconstraint} with,
\begin{align}
   V_{i \bar{i} ; x}{}^y = \sum_k N_{x i}{}^k N_{k \bar{i}}{}^y \; .
\end{align}
We call these defect lines {\em elementary}.  Note that the defect
lines are also in one-to-one correspondence with the primary fields
and so we label them by the same set.  Having obtained this
solution, further solutions may be constructed by 
taking linear combinations with positive integer coefficients.

%We end this section by making two observations.  First, when $x=0$
%one has,
%\begin{align}
%  X_0 = \sum_j P^{(j,\bar{j})} = \text{id} \; , 
%\end{align}
%So the zero disorder line is the insertion of the identity operator.
%Second we observe that in diagonal theories, the number of disorder
%lines of this type, is equal to the number of conformal boundary
%conditions.

\resection{Considerations from TCSA}
\label{sec:t}

The truncated conformal space approach (TCSA) to studying perturbed
quantum field theories is a very powerful tool.  Initially proposed by
Yurov and  
Zamolodchikov \cite{YurZam90}, the method has been successfully applied to
boundary perturbations
\cite{DorPocTatWat98,DorRunTatWat00,GraRunWat00}.   We will begin by reviewing the general
formalism.

Consider a strip of width $L$ with boundary condition $a$ on the
left, $b$ on the right and a perturbation by relevant boundary fields
$\phi_i(x)
=\phi_{i_\alpha ; {\bf n}}^{(bb)}(x)$ applied to the right boundary. 
 We will study this deformation 
through the perturbed Hamiltonian,
\begin{align}
  H_{\text{pert}} = H_0 + \sum_i \lambda_i \phi_i(L) \; .
\end{align}
Mapping this system to the upper half plane (UHP), the Hamiltonian may be
written in terms of operators acting on the UHP Hilbert space,
\begin{align}
  H_{ab}(L) = \frac{\pi}{L} \left( L_0 - \tfrac{c}{24} \right) + \sum_i \lam_i \left(
  \frac{L}{\pi} \right)^{\!\! {-}h_i} \phi_i(1) \; ,
  \label{eq:TCSAham}
\end{align}
where $h_i$ is the conformal weight of the field $\phi_i$.
The idea behind the TCSA is to diagonalise $H_{ab}(L)$ on
a finite dimensional subspace of the unperturbed
Hilbert space.  Such a truncation may be achieved by disregarding all
states above a certain level, $N$.  
This produces an approximation to the finite size system, in
particular to the spectrum and the matrix elements of local fields. The errors
introduced by the cut-off are not particularly well understood, but by
comparison with perturbation theory and exact results for integrable
systems, this cut-off system can approximate the exact system very
well for large ranges of $L$, with errors which reduce on increasing N.

The unperturbed Hilbert space for this system is,
\begin{align}
  \mathcal{H}_{ab} = \bigoplus_k N_{ka}{}^b \mathcal{R}_k \; .
\end{align}
So we denote a basis for the Hilbert space %\footnote{This basis may be
%  different from those used in previous sections.}
 by 
%$\{ | k_\gamma, {\bf n} \rangle \}_{{\bf n}}$ 
$\{ | \phi_{ k_\gamma; {\bf n}} \rangle \}_{{\bf n}}$ 
where $k$ labels the representation, ${\bf n}$ denotes the descendant
states and the label $\gamma = 1,2,\ldots ,N_{ka}{}^b$ encodes any
extra multiplicity there may be.  We chose this basis to be such that
the conformal generator $L_0$ is diagonal.  To simplify notation, we often 
suppress indices labelling basis vectors.

The Hilbert space is equipped with a non-degenerate inner product. 
 In conformal field
theory this factorises into a contribution from the representation
theory of $\mathcal{A}$ and a piece coming from the field theory,
%\begin{align}
%  G_{k,\ell} &= \langle k | \ell \rangle = \langle k_\alpha ,
%  {\bf n} | \ell_\beta , {\bf m} \rangle 
%%  = \lim_{x \to
%%  \infty} x^{2h_k} \langle
%%  \phi_{k_\alpha, {\bf n}}(x) \phi_{\ell_\beta, {\bf
%%  m}}(0) \rangle \notag \\
%%  &=
%  =  g(k)_{{\bf n},{\bf m}} \delta_{k\ell} \langle k_\alpha , {\bf 0}
%  | k_\beta , {\bf 0} \rangle \; ,
%  \label{eq:tcsa2pt}
%\end{align}
\begin{align}
  G_{k,\ell} &= \langle \phi_k | \phi_\ell \rangle = \langle \phi_{ k_\alpha ;
  {\bf n}} | \phi_{\ell_\beta ; {\bf m}} \rangle 
  =  g(k)_{{\bf n},{\bf m}} \, \delta_{k\ell} \, \langle \phi_{ k_\alpha ;
  {\bf 0} } | \phi_{ k_\beta ; {\bf 0} } \rangle \; ,
  \label{eq:tcsa2pt}
\end{align}
where the matrix $g(k)$ is determined entirely from representation
theory.
%,
%\begin{align}
%  g(k)_{{\bf n},{\bf m}} = \langle k,{\bf n} | k, {\bf m} \rangle
%  \left( \langle k,{\bf 0} | k, {\bf 0} \rangle \right)^{-1} \; .
%\end{align}
The matrix elements of the perturbed Hamiltonian are as follows,
\begin{align}
  H|\phi_\ell \rangle = \sum_k H_{k \ell} | \phi_k \rangle \; , \hspace{10mm}
  H_{k\ell} = \sum_j G^{-1}_{kj} \langle \phi_j | H | \phi_\ell \rangle \; .
\end{align}
There are two terms contributing to this.  First the unperturbed
Hamiltonian: in our basis this acts diagonally,
\begin{align}
  \left[ L_0 - \tfrac c{24} \right]_{k \ell} = \delta_{k \ell} \left(
  h_\ell - \tfrac c{24} \right) \; .
\end{align}
The interaction term is given in terms of the three point function:
\begin{align}
  \left[ \phi_{i}(1) \right]_{k\ell} = \sum_j G^{-1}_{kj}
  \langle \phi_j | \phi_{i}(1) | \phi_\ell \rangle \; .
  \label{eq:tcsaInt}
\end{align}
This can be calculated using the chiral algebra to reduce a general
three point function to linear combination of a {\em finite} set of elementary
three point functions multiplied by purely representation theoretic
factors.  Indeed, using the OPE \eqref{eq:OPE} we obtain,
\begin{align}
  \langle \phi_{ j } | \phi_{i}(1) | \phi_{ \ell }
  \rangle = \sum_k \sum_{\gamma=1}^{N_{i\ell}{}^k}
  C_{i\ell ;}{}^k_{\gamma} \, \beta^\gamma \!\! \left[_{i \,
  \ell}^{\; k} \right] G_{j k}
%  \langle \phi_{ j_\alpha ; {\bf n} } | \phi_{i}(1) | \phi_{
%  \ell_\beta ; {\bf m} }
%  \rangle =  F(j,i,\ell)_{{\bf n},{\bf m}} \langle \phi_{ j_\alpha ;
%  {\bf 0} } |
%  \phi_i(1) | \phi_{  \ell_\beta ; {\bf 0} } \rangle \; .
\end{align}
Placing this into \eqref{eq:tcsaInt} we find the two point functions cancel
and so the perturbation matrices are given by representation theoretic data
multiplied by a structure constant.  Replacing all the indices gives
the following formula for the perturbed Hamiltonian,
\begin{align}
   &\left[ H_{ab}(L) \right]_{(k_\alpha,{\bf n }),(\ell_\beta ,{\bf
       m})} 
% \notag \\ & \qquad 
  = % \frac{ \pi }{ L } %\left(  
     \delta_{(k_\alpha,{\bf n }),(\ell_\beta ,{\bf m})} 
    \frac{ \pi }{ L } \left( h_{\ell;{\bf m}} - \tfrac c{24} \right) 
   + \sum_{i, {\bf p},\delta,\gamma} 
    \lambda_{i_\delta;{\bf p}}^{(bb)} \left(
   \frac{L}{\pi} \right)^{\!\! {-}h_{i;{\bf p}}} \! \!
  \bc bba{i_\delta}{\ell_\beta ; \gamma}{k_\alpha}
  \, \beta^\gamma \!\! \left[ _{i ; {\bf p} \;\ell ; {\bf m}}^{\;\;\; k ;
  {\bf n}} \right]
%   C_{i \ell_\beta}{}^{\hspace{-1mm} k_\alpha} 
%     M(k,i,\ell)_{{\bf n},{\bf m}}   \right)
  \; .
  \label{eq:TCSAmat}
\end{align}

%\resection{Postthesis stuff}

\resection{Disorder Lines and Boundary States}
\label{sec:dlandbs}

An important property of the disorder lines is their relationship
with boundary conditions.  Consider a boundary without field
insertions and a disorder 
line running parallel to it.  Looking at the action of a disorder line
on a boundary state we use the 
Verlinde formula \eqref{eq:verlinde} to obtain,
\begin{align}
  X_x | a \rangle &= \left( \sum_j \frac{S_{xj}}{S_{0j}}
  P^{(j,\bar{j})} \right)
  \left( \sum_i \frac{S_{ai}}{\sqrt{S_{0i}}} |i\ir \right) 
  = \sum_j
   \frac{S_{xj}}{S_{0j}} \frac{S_{aj}}{\sqrt{S_{0j}}} |j\ir \notag \\
  &= \sum_{b,j} N_{xa}{}^b \frac{S_{bj}}{\sqrt{S_{0j}}} |j\ir
  = \sum_b N_{xa}{}^b |b \rangle \equiv |x \times a \rangle \; .
\end{align}
And so for example,
\begin{align}
  |a \rangle = X_a |0\rangle \; ,
\end{align}
where here $|0 \rangle$ denotes the boundary state associated to the
identity representation. \\

The partition function of the
strip with Cardy boundary condition ``$a$'' on the left and ``$b$'' on
the right is given in terms of the Verlinde fusion numbers as follows, 
\begin{align}
  Z_{ab} = \sum_i N_{ia}{}^b \chi_i(q) \; .
\end{align}
In particular, one observes that,
\begin{align}
  Z_{0,a\times b} = Z_{\con{a}\times \con{b},0} = Z_{\con{a},b} \; .
\end{align}
These relations become more obvious when one considers a disorder line
being pulled off one boundary, moved across the strip and fused onto
the other.  The question we would now like to address is: 
in what way are the boundary conditions
$a$ and $b$ related to $a\times b$? \\

\subsection*{Embedding Theorem}

\begin{figure}
\begin{center}
\begin{tabular}{cc}
  \refstepcounter{figure}
  \label{fig:defectdrawing} 
  \setlength{\unitlength}{1.25cm}
    \begin{picture}(7,5)
      \put(0,0){ \epsfysize 5\unitlength 
                 \epsfbox{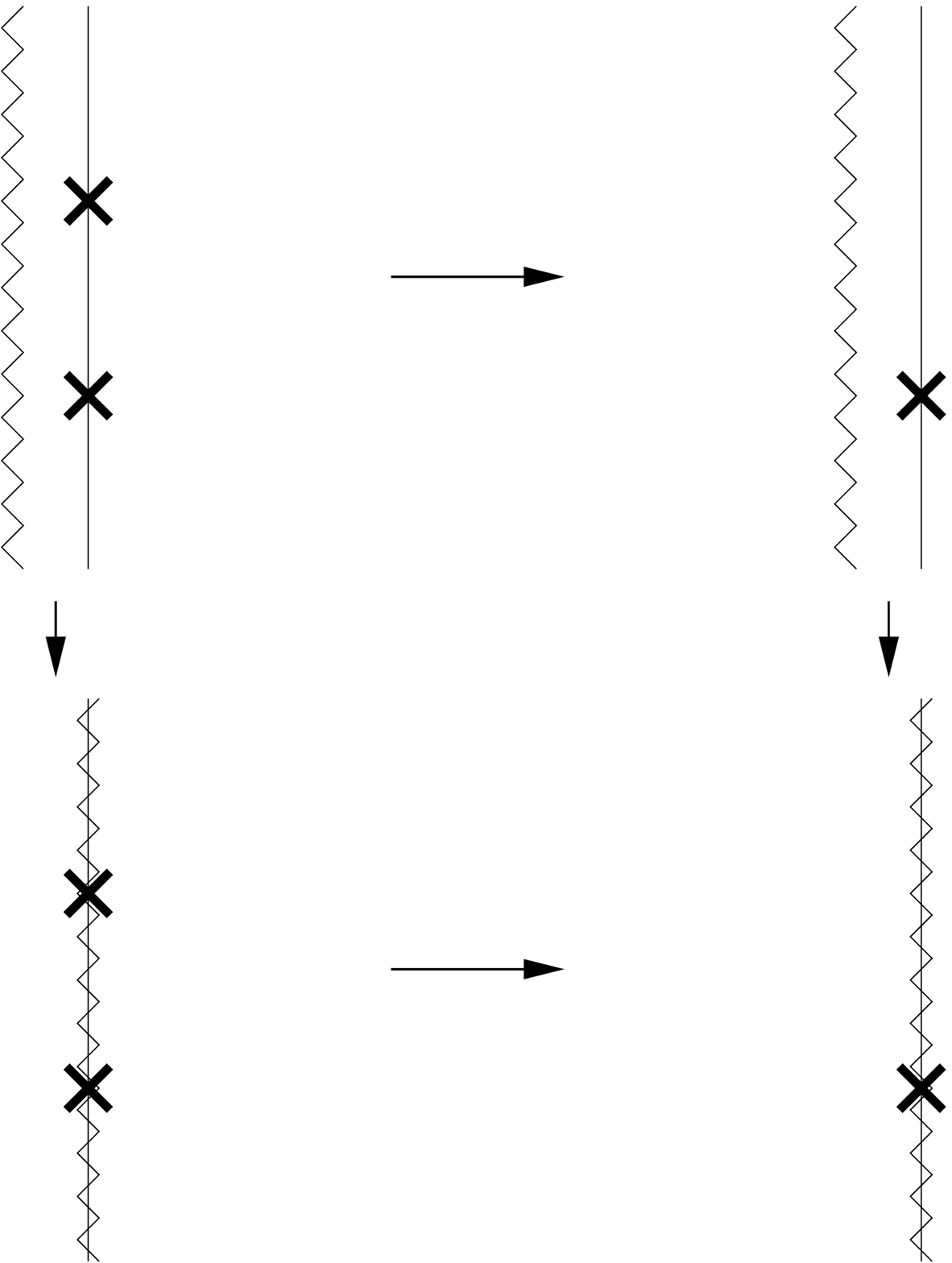}}
      \put(0.7,4.3){\makebox(0,0)[l]{$\phi^{(bb)}_i$}}
      \put(0.7,3.5){\makebox(0,0)[l]{$\phi^{(bb)}_j$}}
      \put(4,3.5){\makebox(0,0)[l]{$\bc bbbijk   \phi^{(bb)}_k$}}
      \put(2.8,3.5){\makebox(0,0)[l]{$\sum_k $}}
      \put(0.7,1.6){\makebox(0,0)[l]{$\phi^{(cc)}_{\sigma(i)}$}}
      \put(0.7,0.7){\makebox(0,0)[l]{$\phi^{(cc)}_{\sigma(j)}$}}
      \put(4,0.7){\makebox(0,0)[l]{$\bc bbbijk \phi^{(cc)}_{\sigma(k)}$}}
      \put(3,0.7){\makebox(0,0)[l]{$\sum_k$}}
      \put(4,0){\makebox(0,0)[l]{$= \sum_n \bc
                 ccc{\sigma(i)}{\sigma(j)}{n} \phi_n^{(cc)} $}}
    \end{picture}
 \\[5mm]
\parbox{9cm}{{\em {\begin{center} figure \ref{fig:defectdrawing} : An illustration
    of the defect line manoeuvres used to obtain the embedding theorem.\end{center}}}} \\
\end{tabular}
\end{center} 
\end{figure}
Consider a boundary with boundary condition $b$, a 
boundary field $\phi^{(bb)}_i(x_i) = \phi_{i_\alpha ; {\bf n}}^{(bb)}(x_i)$ and a defect line $X_a$ running
parallel to it.  The following manipulations are illustrated in figure
\ref{fig:defectdrawing}.  
Applying the defect line to the boundary we see that away from the
insertion points, the boundary condition is changed, $b \to c = a \times
b$.  Applied to the boundary field, the defect line cannot change the
representation theoretic qualities of the field, but it may change
other attributes.  Hence we can write,
\begin{align}
  X_a\left[ \phi_i^{(bb)}(x) \right] = \phi_{\sigma(i)}^{(a\times
  b,a\times b)}(x) = \sum_n D_{in} \phi_{n}^{(a\times
  b,a\times b)}(x)   \; .
  \label{eq:d:embeddingdefn1}
\end{align}
which defines the map $\sigma$ and the numbers $D_{in}$.
Now consider a pair of boundary fields.  Before applying the defect
line, we can use the OPE to simplify the product of the two fields,
\begin{align}
   X_a\left[ \phi_i^{(bb)}(x_i) \phi_j^{(bb)}(x_j)\right] 
  &=  \sum_k \bc bbbijk (x_i-x_j)^{-h_i-h_j+h_k}  X_a\left[
  \phi_k^{(bb)}(x_j)\right] \\
  &=  \sum_{k,n} \bc bbbijk (x_i-x_j)^{-h_i-h_j+h_k} D_{kn} \phi_{n}^{(a\times
  b,a\times b)}(x_j) \; ,
  \label{eq:d:version1}
\end{align}
wherein, to lighten notation we have absorbed the factors of $\beta$
from the OPE \eqref{eq:OPE} into a redefinition of the structure
constants.  
Alternatively, we can apply the defect line before calculating the
OPE.  Let $c=a\times b$ then,
\begin{align}
  X_a\left[ \phi_i^{(bb)}(x_i) \phi_j^{(bb)}(x_j)\right] 
  &= \sum_{n,m} D_{in} D_{jm} \phi_{n}^{(cc)}(x_i)
  \phi_{m}^{(cc)}(x_j) \; \\
  &=  \sum_{n,m,p}  D_{in} D_{jm} \bc ccc{n}{m}p (x_i-x_j)^{-h_i-h_j+h_p}
  \phi_p^{(cc)}(x_j) \; .
  \label{eq:d:version2}
\end{align} 
It
should not matter in which order we apply the defect line or take the
OPE, thus we may identify \eqref{eq:d:version1} and
\eqref{eq:d:version2} and obtain a homomorphism of algebras,
\begin{align}
  \sum_k \bc bbbijk D_{kp} = \sum_{n,m} D_{in} D_{jm} \bc ccc{n}{m}p
  \; .
  \label{eq:d:screlate}
\end{align}
Furthermore, from the explicit form of the defect line operator
(\ref{eq:disorderform},\ref{eq:disordersoln}) we see that applying the
defect line to the identity operator on the $b$ boundary gives a
non-zero result,
\begin{align} 
  X_a \left[ \One^{(bb)} \right] \ne 0 \; ,
\end{align}
and hence the mapping $X_a$ is one-to-one and has a (left) inverse.
Indeed, assume there exists a field $\phi_i^{(bb)}(x)$ such
that $X_a[ \phi_i^{(bb)}(x) ] = 0$, then consider,
\begin{align}
  X_a[ \phi_i^{(bb)}(x) \phi_{\con{i}}^{(bb)}(y) ] = X_a[
  \phi_i^{(bb)}(x)]X_a[ \phi_{\con{i}}^{(bb)}(y) ] 
  \; = \;
  X_a[ (x-y)^{-2h_i} \One^{(bb)} + \ldots ] \ne 0 \; ,
\end{align}
giving a contradiction.
%that applying a defect line to a state is an invertible operation,
%and so the matrix $D_{in}$ must have an inverse.  
%All in all we find that,
%\begin{align}
%  \bc bbbijk = \sum_{n,m,p} D_{in} D_{jm} \bc ccc{n}{m}p D^{-1}_{pk}
%  \; ,
%  \label{eq:d:screlate}
%\end{align}
Hence we have demonstrated an embedding of the operator algebra of the
boundary $b$ into that of the boundary $a \times b$.
Let us formulate these observations in the following theorem, \\

{\em \noindent Let $c=a \times b$ be a relation between labels of boundary
conditions.  Then there exists an isomorphic embedding $b \subset c$ of
the operator algebra of the $b$ boundary condition into that of the
$c$ boundary condition.} \\

\subsection*{Application to Flows}

We now turn our attention to perturbations of the boundary.  Because
the operator algebra of the boundary $b$ can be realised within the
boundary $c= a \times b$, it is natural to ask if the boundary flows
of the boundary $b$ also have a counterpart within the $c$ boundary.
This is indeed the case and in this section we will prove the following theorem:\\

{\em \noindent Let $c = a \times b$ be a relation between labels of boundary
  conditions and let there exist a boundary renormalisation group flow
  from boundary condition $b$ to $d$.  Then there also exists a flow
  from $a \times b$ to $a \times d$.}\\

\noindent Moreover, through our investigations we will obtain
further insight into the embedding theorem.

We begin our proof by considering the matrix elements of the perturbing
Hamiltonian on a pair of strips.  We take strip A to be
$\mathcal{H}_{a^\vee\!,b}$  and strip B to be $\mathcal{H}_{0,a\times b}$.  Note
that the Hilbert spaces of these two systems are identical,
\begin{align}
  \mathcal{H}_{a^\vee\!,b} = \bigoplus_k  N_{ka^\vee}{}^{b}
  \mathcal{R}_k
  = \bigoplus_{k} N_{ab}{}^k \mathcal{R}_k
  = \bigoplus_{k,n} N_{ab}{}^n N_{k0}{}^n \mathcal{R}_k
  = \mathcal{H}_{0,a\times b} \;
\end{align}
and so the unperturbed Hamiltonians are the same.  Here we have used
some of the properties of the fusion numbers 
collected in section \ref{sec:intro}.    
Let us label the vectors of the representation space $\mathcal{R}_k$ by
$|\phi_{ k_\gamma ; {\bf n} } \rangle$ with $\gamma = 1,2,\ldots, N_{ka^\vee}{}^{b}$.
  We will now show
that for any perturbation on strip A, there exists a perturbation of
strip B which has the same perturbing Hamiltonian.  That being so, by
hypothesis we know there exists a flow taking,
\begin{align}
  \mathcal{H}_{a^\vee\!,b} \to \mathcal{H}_{a^\vee\!, d} \; . 
\end{align}
Interpreting this as a flow on strip B we have,
\begin{align}
  \mathcal{H}_{0,a \times b} \to \mathcal{H}_{0,a\times d} \; , 
\end{align}
as required.

We now show that we can find an appropriate Hamiltonian on strip B.
From the embedding theorem, we know that there exists a one-to-one
mapping of each field on the right of strip A into the fields on
the right of strip B preserving
the chiral representation properties of that field.  We also know from
section \ref{sec:t} that
the perturbed parts of the Hamiltonians
are given by a coupling constant times a structure constant and a
representation theoretic matrix,
\begin{align}
  \left[ H_{\text{Pert}}^A \right]_{(k_\alpha,{\bf n}),(\ell_\beta,{\bf
 m})} 
%   &= \sum_{i_\sigma \in \mathcal{H}_{bb}} 
%    \lambda^A_{i_\sigma} \left(
%   \frac{L}{\pi} \right)^{-h_{i_\sigma}} \! \!
%%   C^A_{i_\sigma \ell_\beta}{}^{\hspace{-1mm} k_\alpha} 
%    \bc bb{a^\vee}{i_\sigma}{\ell_\beta}{k_\alpha}
%     M(k,i,\ell)_{{\bf n},{\bf m}} \; .
 &=  \sum_{i, {\bf p},\delta,\gamma} 
    \lambda_{i_\delta;{\bf p}}^{(bb)} \left(
   \frac{L}{\pi} \right)^{\!\! {-}h_{i;{\bf p}}} \! \!
  \bc bb{\con{a}}{i_\delta}{\ell_\beta ; \gamma}{k_\alpha}
  \, \beta^\gamma \!\! \left[ _{i ; {\bf p} \;\ell ; {\bf m}}^{\;\;\; k ;
  {\bf n}} \right]
  \label{eq:d:hamA} 
\end{align}
To write the Hamiltonian of strip B we need a little more notation.
Note that the sectors of the Hilbert space and the Cardy boundary
conditions on the right of the strip 
are labelled by the same set, 
\begin{align}
  a \times b = \sum_k N_{ab}{}^{k} k = \sum_k N_{ka^\vee}{}^b k \; ,
\end{align}
Indeed, 
a boundary field that changes the boundary condition $\ell_\beta$
into $k_\alpha$ also maps the Hilbert space sector 
$\ell_\beta$ into $k_\alpha$ ($\alpha=1,2,\ldots,N_{ab}{}^k$ and
$\beta=1,2,\ldots,N_{ab}{}^{\ell}$) 
.  Hence we can write,
\begin{align}
  \left[ H_{\text{Pert}}^B \right]_{(k_\alpha,{\bf n}),(\ell_\beta,{\bf
 m})}  
%  &= \sum_{i_\rho \in \mathcal{H}_{a\times b,a\times b}} 
%%    \lambda^B_{{(i_\rho,\ell_\beta,k_\alpha)}} 
%    \lambda^B_{i_\rho} 
%  \left( \frac{L}{\pi} \right)^{-h_{i_\rho}} \! \!
%%   C^B_{i_\rho \ell_\beta}{}^{\hspace{-1mm} k_\alpha}
%   \bc {k_\alpha}{\ell_\beta}{0}{i_\rho}{\ell_\beta}{k_\alpha}
%     M(k,i,\ell)_{{\bf n},{\bf m}} \; .
  &= \sum_{i, {\bf p},\rho,\gamma} 
    \lambda_{i_\rho;{\bf p}}^{(k_\alpha \ell_\beta)} \left(
   \frac{L}{\pi} \right)^{\!\! {-}h_{i;{\bf p}}} \! \!
  \bc {k_\alpha}{\ell_\beta}{0}{i_\rho}{\ell_\beta ; \gamma}{k_\alpha}
  \, \beta^\gamma \!\! \left[ _{i ; {\bf p} \;\ell ; {\bf m}}^{\;\;\; k ;
  {\bf n}} \right]
  \label{eq:d:hamB} 
\end{align}
Moreover, using \eqref{eq:structurezero}\footnote{Note that the
  multiplicity labels on $k$ (and $\ell$) in \eqref{eq:d:hamB} range
  over $1,2,\ldots,N_{kb^\vee}{}^c$ and so should not be identified  
  with the multiplicity labels of \eqref{eq:structurezero}.
  Instead, one should take (for example) $[\ell_\beta]_\eps$ with $\eps
  =1,2,\ldots, N_{\ell_\beta 0}{}^{\ell_\beta} = 1$.  Hence we have
  suppressed these indices in \eqref{eq:d:hamB}.}
we find that the structure constant is not equal to one if
and only if the matrix 
%$\beta^\gamma \!\! \left[ _{i ; {\bf p} \;\ell ; {\bf m}}^{\;\;\; k ;
%  {\bf n}} \right]$ 
$\beta^\gamma \!\! \left[ _{i \;\ell}^{\; k} \right]$ 
vanishes: this observation implies that if we set,
\begin{align}
  \lam^{(k_\alpha \ell_\beta)}_{i_\gamma;{\bf p}}
   = \sum_\delta \lam^{(bb)}_{i_\delta;{\bf p}} \bc
  bb{a^\vee}{i_\delta}{\ell_\beta ; \gamma }{k_\alpha} \; ,
  \label{eq:embmap}
\end{align}
we obtain the required 
identification between \eqref{eq:d:hamA} and \eqref{eq:d:hamB}. \\

Actually to obtain \eqref{eq:embmap} and prove our main result, we
have not used the embedding theorem at all.  Instead, equation
\eqref{eq:embmap} provides a candidate 
formula for the matrix $D_{in}$ defined in equation
\eqref{eq:d:embeddingdefn1}:  let $k_\alpha, \ell_\beta$ label the elementary
boundaries comprising $c = a \times b$ as above then,    
\begin{align}
  X_a \left[ \phi^{(bb)}_{i_\delta ; {\bf p}} \right] =
  \sum_{k_\alpha , \ell_\beta , \gamma} 
  \bc bb{a^\vee}{i_\delta}{\ell_\beta ; \gamma }{k_\alpha}
  \phi^{(k_\alpha \ell_\beta)}_{i_\gamma ; {\bf p}}
  \; ,
  \label{eq:embed2}
\end{align}
and equation \eqref{eq:d:screlate} can be seen by a straightforward
application of 
the pentagon identity for the fusing matrix \cite{MooSei89} (also
\cite{FelFroFucSch99b}).  We have not shown that the embedding
\eqref{eq:embed2} is the same as that obtained using a defect
line \eqref{eq:d:embeddingdefn1}, however it seems very likely.  
One way to show this would be to use the 3-dimensional topological
realisation of defect lines and boundary conditions introduced in
\cite{FucRunSch02}.

\resection{Applications}
\label{sec:apps}

In this section, we apply our rule to a number of renormalisation group
flows studied in the literature.

\subsection{Minimal Model Flows}
\label{sec:app:mm}

In this subsection we consider the A-series unitary minimal models
$M(m,m{+}1)$ with central charge,
\begin{align}
  c=1-\tfrac{6}{m(m+1)} \; .
\end{align}
The set labelling representations, primary bulk fields, boundary conditions and
defect lines is the Kac table $K$ where,
\begin{align}
  K' &= \{ (r,s) \; : \; 1 \le r \le m{-}1 \; , \; \; 1 \le s \le m \}
  \; , \\
  K &= K' / \sim \; \qquad \text{where } \; \; (r,s) \sim
  (m{-}r,m{+}1{-}s) \; .
\end{align}
Let us now consider the boundary renormalisation group flows in these
models.  In \cite{RecRogSch00}, Recknagel {\em 
el.~al.~} studied  boundary flows in the unitary minimal models using
perturbation theory.  There it was found that on perturbing the
boundary condition $(r,s)$
by the field $\phi_{13}$ in the limit of large $m$, one flows to the
superposition of boundary conditions, 
\begin{align}
  (r,s) + \phi_{13} \to \oplus_{i=1}^{\min\{r,s\}} (|r{-}s|{+}2i{-}1,1) \; ,
\end{align}
Here we will extend this result to finite $m$ and prove this flow
exists for $s \le \tfrac{m+1}{2}$.  Our result depends on the work of
\cite{LesSalSim98} showing that the following flow exists,
\begin{align}
  (1,s) + \phi_{13} \to (s,1) \; , \qquad s \le \tfrac{m+1}{2} \; . 
  \label{eq:provenflow}
\end{align}
Indeed, by using our theorem we may identify the flows in $(1,s)$ with
a subset of the flows in $(r,1) \times (1,s) \simeq (r,s)$ as follows,
\begin{align}
%  (r,s) \simeq (r,1) \times (1,s) 
%  \to 
%  (r,1) \times (s,1) \simeq \bigoplus_{\begin{subarray}{c} i=1+|r-s| \\ i+r+s
%  \; \text{odd} \end{subarray}}^{\min\{r+s-1,2m-1-r-s\}}  (i,1) \; . \\ 
  (r,s) \simeq (r,1) \times (1,s) 
  \; \to \; 
  (r,1) \times (s,1) \simeq \oplus_{i=1}^{\min\{r,s,m{-}r,m{-}s\}}
  (|r{-}s|{+}2i{-}1,1) \; . 
  \label{eq:RRSp}
\end{align}
To study the region $s>\tfrac{m+1}{2}$, we can use the identification in the
minimal model $(r,s)\sim(m{-}r,m{+}1{-}s)$ to find \cite{LesSalSim98},
\begin{align}
  (1,s) + \phi_{13} \to (s{-}1,1) \; , \qquad s > \tfrac{m+1}{2} \; . 
  \label{eq:provenflowtoo}
\end{align}
hence,
\begin{align}
  (r,s) &\simeq (r,1) \times (1,s) 
  \; \to \;
  (r,1) \times (s{-}1,1) \simeq
  \oplus_{i=1}^{\min\{r,s{-}1,m{-}r,m{+}1{-}s\}}
  (|r{-}s{+}1|{+}2i{-}1,1) \;  . 
  \label{eq:RRSm}
\end{align}
%In agreement with the numerical studies
%\cite{Wat:un} (see \cite{RecRogSch00}).  It is widely
%believed that the flow \eqref{eq:provenflow} is actually true for
%$s<m$ and \eqref{eq:provenflowtoo} is true for $s>1$.  While there is
%numerical evidence for this through TCSA, we do not believe it has
%been explicitly demonstrated.
%\comment{to Gerard: Can we use \cite{LesSalSim98} as proof of
%  the existence of \eqref{eq:provenflow} and
%  \eqref{eq:provenflowtoo} over the whole range of $s$?  Is what I
%  have said here true?}
It is generally believed that the flows \eqref{eq:RRSp} and
\eqref{eq:RRSm} exist for all $(r,s)$ with $1<s<m$, and are generated
by the same perturbation but with opposite sign of the coupling
constant.  This is borne out by numerical studies \cite{Wat:un} (see
\cite{RecRogSch00}), TBA studies in the case of $r=1$
\cite{LesSalSim98} and is in agreement with the structure of boundary
weights in lattice models \cite{BehPea00,GraWat:ta}.  \\

In \cite{Gra02} one of us studied 
perturbations of superpositions of boundary conditions of the form,
\begin{align}
  \omega = \oplus_{i=1}^{n} (r,s+2i-2) \; .
\end{align}
Note that this superposition may be written,
\begin{align}
  \omega \simeq (r,s{+}n{-}1 ) \times (1,n ) 
         \simeq (r,n) \times (1,s{+}n{-}1 ) \; .
\end{align}
Applying our theorem we reproduce and generalise the flows found in
\cite{Gra02}, 
\begin{align}
  \omega \to 
  \begin{cases}
    (r,s{+}n{-}1 ) \times (n,1 ) 
    \simeq 
    \oplus_{i=1}^{\min{\{ r,n,m{-}r,m{-}n \}}}
    (|r{-}n|{+}2i{-}1 , s{+}n{-}1 ) \; , \\ 
    (r,n) \times (s{+}n{-}1,1 ) 
    \simeq 
    \oplus_{i=1}^{\min{\{ r,s{+}n{-}1,m{-}r,m{-}s{-}n{+}1 \}}} 
    (|r{-}s{-}n{+}1|{+}2i{-}1, n) \; .
  \end{cases}
\end{align}
Furthermore, we can use the non-perturbative flow \eqref{eq:provenflowtoo} to obtain,
\begin{align}
  \omega \to 
  \begin{cases}
    (r,s{+}n{-}1 ) \times (n-1,1 ) 
    \simeq 
    \oplus_{i=1}^{\min{\{ r,n-1,m{-}r,m{-}n{+}1 \}}}
    (|r{-}n{+}1|{+}2i{-}1 , s{+}n{-}1 ) \; , \\ 
    (r,n) \times (s{+}n{-}2,1 ) 
    \simeq 
    \oplus_{i=1}^{\min{\{ r,s{+}n{-}2,m{-}r,m{-}s{-}n{+}2 \}}} 
    (|r{-}s{-}n{+}2|{+}2i{-}1, n) \; .
  \end{cases}
\end{align}
These have also been observed using the absorption of boundary spin
conjecture of \cite{FreSch02a,Fre03}.

\subsection{Coset Models}
\label{sec:app:coset}

As a second application of this analysis, we will demonstrate that the
``absorption of boundary spin'' proposal of Fredenhagen and Schomerus
\cite{FreSch02a,Fre03} may be reduced to a statement about more elementary
flows in some cases.  First we remind the reader of the proposal of
Fredenhagen and Schomerus.  The proposal concerns boundary flows in
diagonal WZW coset models  
%(see appendix \ref{app:coset} for our notation) 
whose field
content is specified by two affine Lie algebras $\mathfrak{h}_{k'}
\subset \mathfrak{g}_k$.  A set of elementary boundary conditions for
this theory 
may be labelled by pairs of integrable representations of these
algebras, $(L,L')$ where $L$ is an irreducible representation  of
$\mathfrak{g}_k$ 
and $L'$ is an irreducible representation of $\mathfrak{h}_{k'}$.
However, not all pairs allowed and some pairs label the same boundary
condition.  

The selection and identification rules are specified in terms of two objects\footnote{We
refer the reader to \cite{Fre03,DiFMatSen97} for details.}: the
{\em identification group} $G_{\text{id}}$, a subgroup of the
set of simple currents in $\mathfrak{g}_k {\otimes} \mathfrak{h}_{k'}$
and which we assume to act fixed point free\footnote{Cosets can be
  defined in more general situations, however like
  \cite{FreSch02a,Fre03}, we will not consider with such models.}; and the
{\em monodromy charge} associated to a simple current $J$, $Q_J(L) =
h_J + h_L - h_{J {\times} L} \;  \text{mod}  \; \mZ$,  where $h_L$ is
the conformal weight of the representation $L$ in either
$\mathfrak{g}_k$ or $\mathfrak{h}_{k'}$ as appropriate.
\begin{itemize}
\item A pair $(L,L')$ is allowed if $Q_J(L) = Q_{J'}(L')$ for all $(J,J')
\in G_{\text{id}}$. 
\item The pairs $(L,L')$ and $(J {\times} L, J' {\times} L')$ are
identified for each $(J,J') \in G_{\text{id}}$.
\end{itemize}
A general boundary condition in the coset theory is a superposition of
elementary boundary conditions so one may also associate a boundary
condition to a pair of representations $(L,L')$ by expanding in terms
of irreducibles.
We may now state
the absorption of boundary spin conjecture: \\

{\em \noindent Let $(S,0)$ and $(L,L')$ be representations in
  $\mathfrak{g}_k {\otimes} \mathfrak{h}_{k'}$ such that
%  $Q_{(J,\con{J'})}( S,0) + Q_{(J,\con{J'})}(L,L') = 0$ for all
%  $(J,J') \in G_{\text{id}}$ 
  $Q_J(S) + Q_J(L) = Q_{J'}(L')$ for all
  $(J,J') \in G_{\text{id}}$,  
then a flow exists between the following pair of boundary conditions,
\begin{align}
  (L,\con{S} |_{\mathfrak{h}}\times L') \to (L \times S , L')
  \label{eq:absrule}
\end{align}
where the restriction on the left hand side involves taking $\sigma_0$
the corresponding 
representation of the finite dimensional Lie algebra $\mathfrak{g}$,
and restricting this to the subalgebra $\mathfrak{h} \subset
\mathfrak{g}$.} \\

To use our result, we need to factorise \eqref{eq:absrule} into pairs
of representations allowed by the coset selection rules.  This can be
achieved if $Q_J(S)=0$, in which case we use the fusion rules of the coset
theory to rewrite \eqref{eq:absrule} as follows,
\begin{align}
  (L,L') \times (0,\con{S}|_{\mathfrak{h}}) \to (L,L') \times (S , 0) \; .
\end{align}
In this form it is trivial to see using the
machinery developed in this paper that the existence of the
flow,
\begin{align}
  (0,\con{S} |_{\mathfrak{h}}) \to (S , 0) \; ,
\end{align}
implies the existence of the others.  Thus we have reduced a
particular case of the absorption of
boundary spin proposal of Fredenhagen and Schomerus to a statement
about simpler flows.  It remains to be shown that these elementary
flows actually exist.

\resection{Conclusions}
\label{sec:conc}

In this note we have used the technology of defect lines to
demonstrate an embedding of the operator algebra of a Cardy boundary
condition $b$ into that of $a \times b$.  We then used this
observation to show that if there exists a flow between boundary
conditions $b$ and $d$ then there also exists a flow between $a \times
b$ and $a \times d$ where $\times$ denotes the fusion product.
Finally we considered applications of this rule to the Virasoro
minimal models and to the absorption of boundary spin conjecture of
Fredenhagen and Schomerus \cite{FreSch02a,Fre03}.

One obvious question is in what way does the rule extend to more general
non-diagonal rational conformal field theories?
In this case the interrelationship between the various positive
integer valued matrices discussed in \cite{BehPeaPetZub99,PetZub01} will
be relevant.  A formalism in which both defect lines
and boundary conditions are particularly transparent is the
3-dimensional topological quantum field theory framework championed in
\cite{FucRunSch02,FucRunSch03}. 

In arguing our theorem, we used the matrix elements of the perturbed
Hamiltonian in the unperturbed basis.  We avoided questions of
renormalisation by assuming these problems were solved.  It would be
nice to have a better understanding of the 
renormalisation procedure using level truncation.  Such knowledge may
lead to a new rigorous analytic method for studying
renormalisation group flows.

\subsection*{Acknowledgments}

We would like to thank Roger Behrend, Stefan Fredenhagen, Ingo Runkel
and Volker Schomerus for useful discussions.  
GMTW would also like to thank Z.~Bajnok, L.~Palla and G.~Takacs for
useful discussions of boundary conditions, and the Royal Society for a
joint project grant supporting these discussions.  
This work is supported
in part by the EC network grant ``EUCLID'', contract number
HPRN-CT-2002-00325.

\bibliographystyle{utcaps}%{gerabbrv}
\bibliography{references}

\end{document}